# QUALITY SCALABLE QUANTIZATION METHODOLOGY FOR DEEP LEARNING ON EDGE

Salman Abdul Khaliq and Dr. Rehan Hafiz

***Abstract*—Deep Learning Architectures employ heavy computations and bulk of the computational energy is taken up by the convolution operations in the Convolutional Neural Networks. The objective of our proposed work is to reduce the energy consumption and size of CNN for using machine learning techniques in edge computing on ubiquitous computing devices. We propose Systematic Quality Scalable Design Methodology consisting of Quality Scalable Quantization on a higher abstraction level and Quality Scalable Multipliers at lower abstraction level. The first component consists of parameter compression where we approximate representation of values in filters of deep learning models by encoding in 3 bits. A shift and scale based on-chip decoding hardware is proposed which can decode these 3-bit representations to recover approximate filter values. The size of the DNN model is reduced this way and can be sent over a communication channel to be decoded on the edge computing devices. This way power is reduced by limiting data bits by approximation. In the second component we propose a quality scalable multiplier which reduces the number of partial products by converting numbers in canonic sign digit representations and further approximating the number by reducing least significant bits. These quantized CNNs provide almost same ac-curacy as network with original weights with little or no fine-tuning. The hardware for the adaptive multipliers utilize gate clocking for reducing energy consumption during multiplications. The proposed methodology greatly reduces the memory and power requirements of DNN models making it a feasible approach to deploy Deep Learning on edge computing. The experiments done on LeNet and ConvNets show an increase upto 6% of zeros and memory savings upto 82.4919% while keeping the accuracy near the state of the art.**

***Index Terms*—Quality Scalable, Quantization, Edge Computing, Approximate Computing**

## I. INTRODUCTION

THIS Machine Learning techniques such as “Deep Learning” [1] have enabled the most remarkable innovations in artificial intelligence. Deep Neural Networks find their usage in novel areas such as human activity recognition, image to text translation and autonomous driving cars [1]. However, these impressive methods are much costly in terms of computations. Training a modern DNN architecture is essentially an exascale computational task where a high performance compute system is required. Training of these networks implemented using the cloud or a cluster of high performance GPUs that can consume power varying from hundreds to thousands of megawatts [2]. On the other hand trained DNNs are deployed in power-constrained embedded platforms for using in different applications and feasibility of running deep learning networks on edge is also under consideration.

The energy efficiency is equally important for training using high performance computing systems as much as it is for deploying deep learning on mobile platforms or embedded devices. The architectural strategies for increasing resource utilization and reducing off-chip memory accesses include usage of systolic array [10], parallel computing architecture [11], tiling etc. and algorithmic improvements include pruning [12], quantization [13], efficient gradient descent [14] etc. to reduce the overall size of the deep learning models and speeding computations. Large deep learning models require large memory for model storage and high memory bandwidth for computations of the convolution operations. Figure 1 [8] shows the comparison of energy consumed by add and multiply operations and the energy required for accessing data from the DRAM.

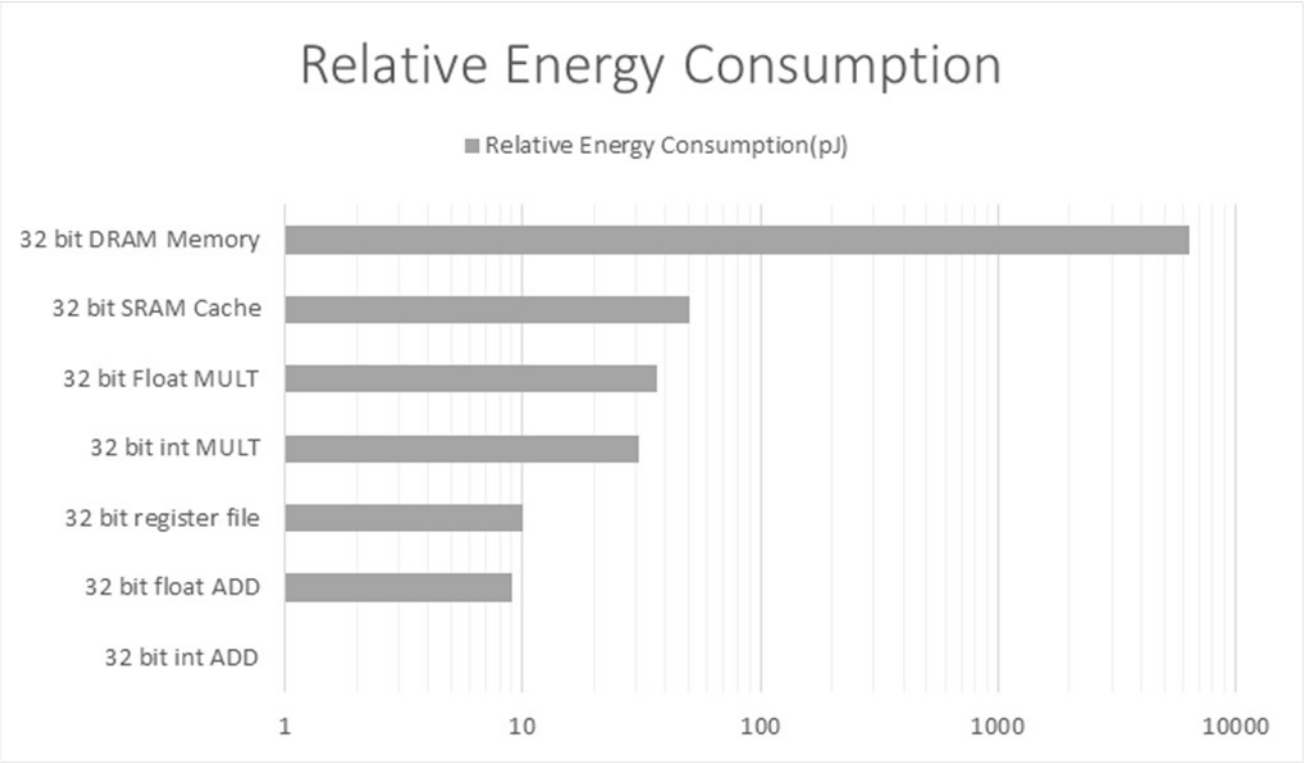

Fig. 1 Energy Consumption comparison between different operations

Memory utilization is therefore an important factor for computing devices. The off-chip memory accesses effect the overall power requirements along with the computational energy consumed by resources. Fig. 2 below shows the components which contribute towards total energy consumed by Neural Network computation system [8].

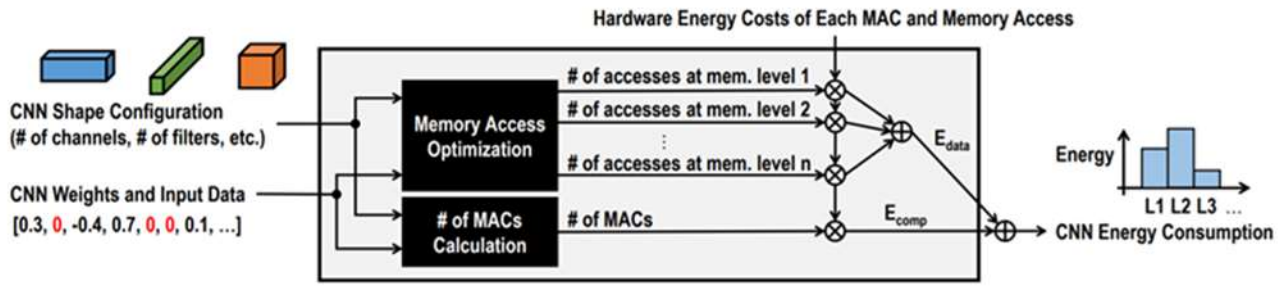

Fig. 2 Contribution in Energy Consumption cost in CNN

The Figure 1 shows that the energy consumed by the memory accesses for retrieving weights and data also contribute to the overall energy taken by CNN networks and Figure 2 shows that it is a major portion of energy requirements. The zero data points are indicated by red because multiplication for these points can be skipped for energy savings.

Some of state of the art architectures optimized for low power in Deep Learning applications include Efficient Inference Engine (EIE) for compressed Deep Neural Network [15], Eyeriss: An efficient and reconfigurable accelerator for deep convolutional neural networks [16] and a machine learning architecture which focuses on small memory footprint and high throughput [11]. These accelerator designs focus on maximizing resource utilization by reducing off-chip memory accesses and data reusability. Hardware technique of zero-skipping is also employed by some low power architecture designs such as EIE after encoding the non-zero values of DNN filters.

The algorithm level optimization for low power deployment of DNN are targeted towards generic hardware platforms and not necessarily for GPU or FPGA platforms. The state-of-the-art algorithms are targeted for low power and model compression include Deep Compression [43] which combines compression techniques of pruning [8], quantization [22], hashing and weight codebook methods to achieve reduced model size. Quantization techniques such as in Binary Neural Networks, XNOR-Net [22] and Ternary Neural Networks [29] are also proposed to reduce the number of computations as well as reducing the model size of DNN.

The concept of edge computing [18] is becoming prevalent for using in different application areas especially deep learning on edge. The basics of edge computing state that the computing devices at the end of communication channel process the information transmitted by the source, in case of deep learning the size of the information transmitted by the source can be reduced by encoding the model before transmitting it over the channel where the computing device at the end decodes the model to be deployed for inference. The edge computing devices have varying computing power which demands the need for quality scalable design. The figure below shows the different computing platforms with widely varying compute resources.

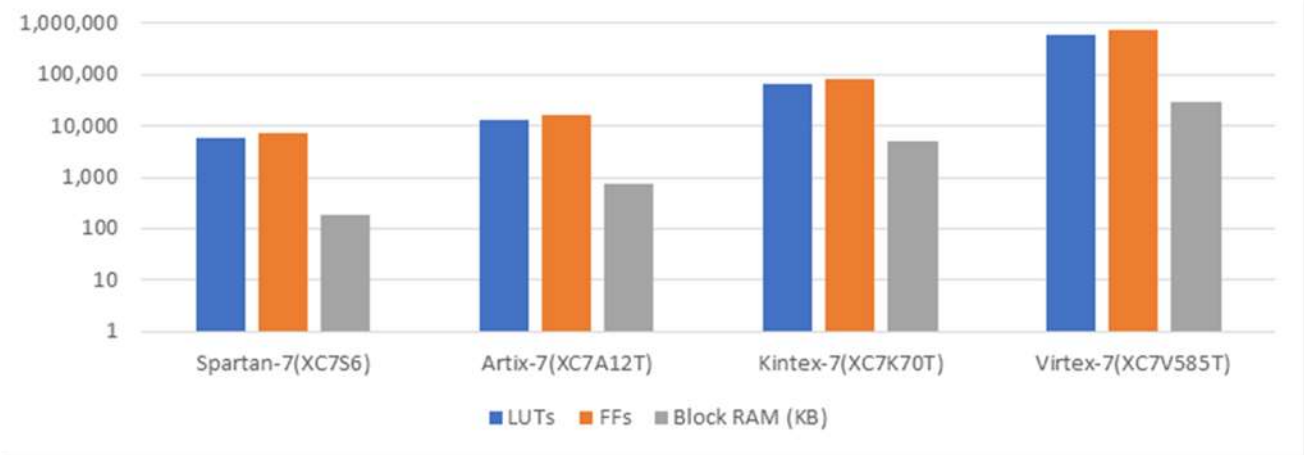

Fig. 3 Resource comparison between different FPGA devices

The mentioned algorithm level and hardware optimization techniques and architectures may prove to be feasible choices to deploy for edge computing on portable devices. However a quality scalable design methodology is required for deploying DNNs on devices with widely varying compute resources on edge. In this paper we propose a systematic methodology for quality scalable design, quality scalable quantization scheme and a Quality Scalable Multiplier design.

## II. LITERATURE SURVEY

In this section we have described some of state of the art algorithm level optimization strategies which we have surveyed. Table 1 below shows the various state of the art algorithm level optimization techniques. The two major techniques aim to reduce the computational cost and compress the DNN model size. The classification of these techniques is made on the basis of how the bit-width is reduced which represents the classification filter values.

The DNN models are able to classify the input data into various classes after training the values in filters of convolution layers. The filters in convolution layers may have many number of weight matrices where each matrix may have many channels known as depth depending upon the output of the previous layer. Weighted quantization techniques are a set of methods which aim to minimize following optimization program

$$J(B,\alpha) = min||W - \alpha B||^2 \quad (1)$$

This is a generic optimization for achieving quantization. Here $\alpha$ represents a scaling number which can be any real number.

TABLE I
CLASSIFICATION OF DIFFERENT TECHNIQUES TO REDUCE BIT-WIDTH OF PARAMETERS IN CNN

| | Technique Employed | Network Architecture and Datasets tested on |
|---|---|---|
| Weighted Quantization | Binary Networks (BNN, BC, XNOR-Net, BWN) | Binary Neural Networks (BNN) [20] Dataset tested on: -MNIST - SVHN - Cifar-10 |
| | | Binary Connect (BC) [21] Dataset tested on: - MNIST - SVHN - Cifar-10 |
| | | XNOR-Net [22] Dataset tested on: - ImageNet using AlexNet |
| | Ternary Weight Networks (TWN) [29], [30] | TWN Dataset Tested on: - MNIST on Lenet-5 - ImageNet on Resnet18B - Cifar 10 on VGG 7 |
| Low-Bit Precision | Increased precision for accuracy (extremely low bit precision) [14] | Neural Networks tested on - AlexNet - VGG16 - ResNet 18 - Resnet 50 |
| | Dynamic Low precision [35] | Datasets used - Lenet - Alexnet |

In case of XNOR-Net [22] $B$ represents a vector such that $B \in \{1,-1\}^n$, $n$ here is the length equal to the number of values in $W$. $W$ represents a single dimensional vector containing values of convolution filters. This optimization program solution for these parameters are computed as follows

$$B^* = sign(W) \quad (2)$$

$$\alpha^* = \frac{1}{n}||W||_{l1} \quad (3)$$

As an extension to the above mentioned Binary Neural Networks *Mellempudi et al* in [45] have proposed a ternary weight network which utilizes the pre-trained weights for quantization. In this case binary vector is given by $B_i \in \{+1,\ 0,\ -1\}$. The thresholds are defined for positive and negative values for $B$ in [45] are computed using following solution

$$\Delta^* = \underset{\Delta>0}{\operatorname{argmax}} \frac{(\sum_{i \in I_\Delta} |W_i|)^2}{|I_\Delta|} \quad (4)$$

However Mellempudi et al in [45] also writes that the brute force method for defining thresholds yields better accuracies for the neural network. Leng et al in [14] have proposed quantization of weights such that there are levels of quantization. These quantization levels are defined as $W \in C \in \{0, \pm\alpha, \pm 2\alpha, \pm 2^2\alpha, \ldots, \pm 2^N\alpha\}$ and $\alpha > 0$. Their idea is derived from the recent development of Alternating Direction Multiplication Method (ADMM) for mixed integer optimization programs. Many mathematical optimizations for the training of these binary and increased discrete quantization have been proposed [46], [47]. These optimizations yield better results for binary and increased quantization techniques in terms of accuracy.

However all the above mentioned techniques reduce the model size and memory requirements, a systematic quality scalable approach is a pressing demand for ubiquitous computing platforms in edge computing. We therefore propose a quality scalable design methodology which could reduce the model size and provide quality scalable accuracy on cost of increased bit-width required to store quantized values of DNN filters.

## III. Proposed Methodology

In the previous chapter we discussed how state of the art techniques provided energy efficient deep learning methods. The algorithm level techniques introduced pruning, binary and ternary quantization and dynamic bit widths which reduce model parameters and computations. These techniques provided energy optimization but quality scalable techniques were needed to configure energy requirements, so that various architectures with different computational resources could be used for deep learning. This quality scalability is also useful for edge computing environments where user end devices vary in compute capability. In this chapter we provide framework for quality scalable methodology for deep learning which incorporate scalable quantization.

### *A. Framework for Quality Scalable Methodology for Deep Learning Models*

In this framework we first of all choose a deep learning model which we need to quantize to achieve parameter reduction. The training of the model is carried out using *Keras* library for deep learning, after the training is complete we extract the trained filter weights to quantize. The trained weights are divided into vectors of length N, for each vector we assume a Gaussian distribution for which we compute mean, and a relative scalar α. We also compute standard deviation for positive and negative vector values represented by $\sigma_P$ and $\sigma_N$ respectively. Then we quantize the weights for different quantization levels, for this experiment we considered three quantization levels. After quantization we fine-tuned the model if required. Then the quality scalable multipliers are employed for convolution operations of scaled quantized weights with input data in forward path. The blocks with blue background is our contribution and white blocks represent tools which are available. Cylindrical blocks represent a container for values, rhombus represents decision and oval blocks represent operations.

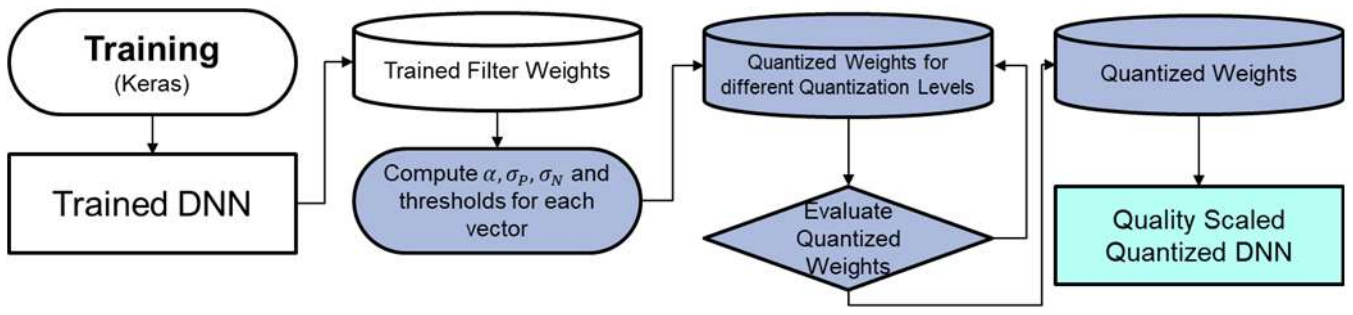

Fig. 4 Framework for the proposed methodology

At each convolution layer. The options available for quantization of values in the filter are analyzed. The available points in design space which gives lowest error in accuracy and minimum memory footprint are determined. The shape of the distribution of trained weight values is used to determine the variance using Maximum Likelihood Estimation.

The thresholds and configuration of minimum Quantization Levels are then determined by exhaustive search. After determination of hyper parameters trained weights are converted into quantized values. The accuracy of the DNN architecture is computed.

### *B. Determining Weights for different Quantization Levels*

Quality Scalable Quantization introduces different quantization levels for better representation of filter values of pre-trained networks. We assume a filter W with dimensions $b \times h \times c$, where b, h and c represent breadth, height and number of channels respectively. Let i be the number of filters with depth of c channels. From here Filter matrix of each channel has b breadth and h height, for each value located at $w \times h$ across all channels are formed in a vector we call $W_i$. We need to find the closest $l_2$ displacement between quantized and original weights, which can be mathematically represented by equation 3.1. Here the optimization program for minimizing error between quantized and original trained weights is given

$$J(\beta, \alpha) = min\left|\left|\overline{W}_i - \alpha_i \overline{\beta_i}\right|\right|^2 \quad (5)$$

Here $\alpha_i$ represents the scalar value associated with each vector and $\overline{\beta}_i$ represents a vector containing quantization values exponentially increasing with powers of 2 such that

$$\overline{\beta_i} \in \{+4, +2, +1, 0, -1, -2, -4\}^c \quad (6)$$

where c are the number of channels in a convolution layer. The variance of the data in the filter channels vary from one filter to another and we estimate the variance and mean of the vector using the "Maximum Likelihood Estimation" (MLE). The shape of PDF of the vector are assumed to be Guassian. The variance of the data when we assume the distribution to be Guassian can be computed empirically using

$$\frac{\sum_{i=1}^{N}(W_i - \overline{W})^2}{N} \qquad (7)$$

The formulation for obtaining number of quantization levels in $\beta$ is given in equation 8. Here $\emptyset$ represents value in format of $2^k$, where k can be any positive integer. $\theta$ gives number of

$$\theta = \left\lceil [log_2(2 \times (1 + log_2(\emptyset))] + 1) \right\rceil \qquad (8)$$

The number $\emptyset$ is used to determine a scaling number $\alpha$ to correspond with the quantization levels in vector $\beta$ using the following equation 9.

$$\alpha_i^* = \frac{\sum_{j=1}^{n} |W_j^i|}{\emptyset \times N} \qquad (9)$$

The above equation is the standard formula for computing statistical mean for a Gaussian distribution except for scaling factor $\theta$. W represents the values in the vector and $N$ is equals length of vector in that convolution layer. This scalar is multiplied with the quantization vector $\overline{\beta_i}$. The quantization value with which the scalar multiplies to give closest approximation is chosen according to difference of the actual value from standard deviation of vector.

$$\beta_i^* = \begin{cases} -4 \; for \; W_i < -\delta \times \sigma_N \\ -2 \; for \; -\delta \times \sigma_N < W_i < -\sigma_N \\ -1 \; for -1 \times \sigma_N < W_i < -\gamma \\ 0 \; for \; -\gamma < W_i < \gamma \\ +1 \; for \; \delta < W_i < 1 \times \sigma_P \\ +2 \; for \; \sigma_P < W_i < \delta \times \sigma_P \\ +4 \; for \; \delta \times \sigma_P < W_i \end{cases} \qquad (10)$$

Here $\sigma_N$ is the standard deviation of the vector containing negative filter values across all channels. $\delta$ and $\gamma$ represent parameters for thresholds of $Q$ values and 0 respectively. When designing an accelerator, the way in which computations are carried out may benefit by choosing quantized vectors either filter wise or channel wise as shown in Figure 5 and Figure 6.

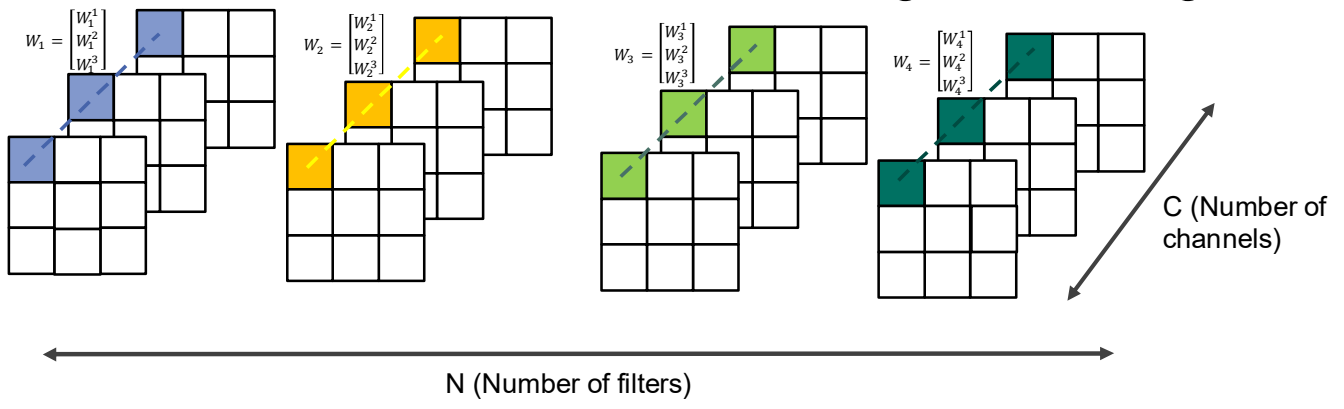

Fig. 5 Selecting channel wise vector from convolution filters

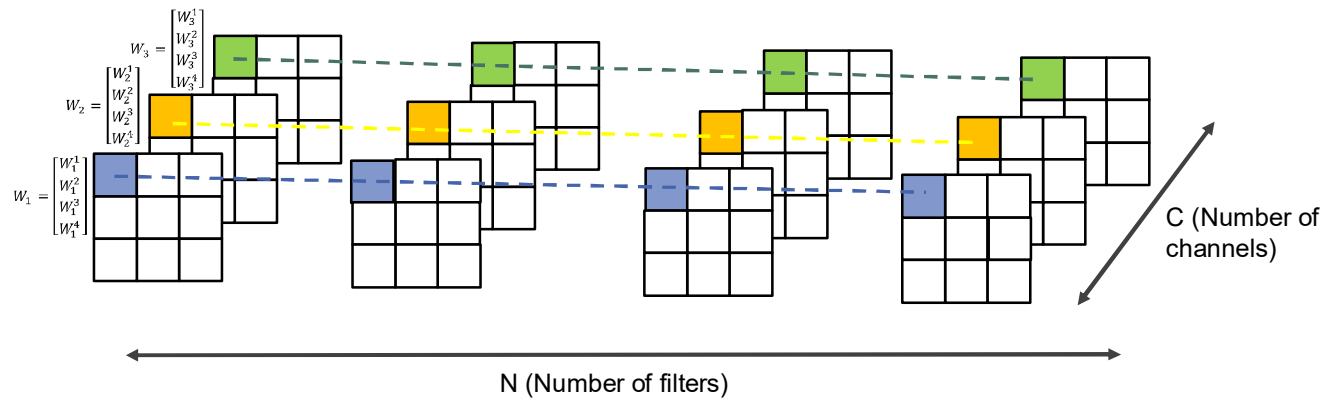

Fig. 6 Selecting filter wise vector from convolution filters

Using the above methodology we can encode the DNN model in a compressed form. We assign values across the channels in a filter with a 3 bit code along with a full precision scalar value. After DNN model is transmitted through a communication channel to the edge computing device, the values are decoded according to the following table. The memory bandwidth in this way is more efficiently utilized and approximate weights are recovered by using only shifting and inverting operations on the scalar number. The table for converting the 3-bit code into approximate filter values is shown below

TABLE II
RECOVERING APPROXIMATE VALUES FROM 3-BIT CODE

| Sr. No | 3-bit Code | Operation |
|---|---|---|
| 1. | 000 (0) | 0 is skipped |
| 2. | 001 (1) | The scalar is used without shifting |
| 3. | 010 (2) | scalar is used after shifting left once |
| 4. | 011 (4) | Scalar is used after shifting left twice |
| 5. | 100 (-1) | Scalar is used after inverting |
| 6. | 101 (-2) | Scalar is used after inverting and shifting right once |
| 7. | 110 (-4) | Scalar is used after shifting right twice and inverting |
| 8. | 111 (no operation) | XX |

## IV. Experimental Results

We tested our methodology on LeNet and ConvNet architectures on MNIST and Cifar-10 datasets respectively using three different configuration modes. We used Keras libraries for training and inference for quantized and normal DNN models. We have performed various experiments to determine the feasibility of quantization and demonstrated quality scalability for CNN models.

### A. *Effect of Quantization on Classification Accuracy*

The classification accuracy of LeNet Architecture on MNIST came out to be 98.86% when trained from scratch. The Quantization was performed on the LeNet Architecture, and without any fine-tuning we achieved an accuracy of 97.59% which is a good enough approximation. The results on LeNet Architecture on MNIST dataset are summarized in Table 3 below. The fine-tuning of FC layer only increases the accuracy upto 98.35%.

TABLE III
ACCURACY ON LENET AFTER QUANTIZATION AND FINE TUNING FULLY CONNECTED LAYER

| LeNet Architecture | Accuracy |
|---|---|
| Without quantizing the weights(30 epochs) | 98.68% |
| After weights quantization(without retraining) | 97.59% |
| After weights quantization(5 epochs only FC) | 98.35% |
| After weights quantization(20 epochs only FC) | 98.55% |

The parameters of LeNet model were reduced upto 82.4919% and quantization also resulted in increased the number of zeros upto 6% as compared to the non-quantized trained model of LeNet.

### B. *Evaluation of Quality Scalable Methodology*

We have evaluated quality scalability by testing three different quantization levels. The weights are quantized by changing the parameter $\emptyset$. We have tested for $\emptyset$ values of 1,2 and 4 and corresponding values of $\alpha$,$\beta$ and $\sigma$. We evaluated two deep learning architectures, LeNet and 4 layer ConvNet on MNIST hand written number recognition and Cifar-10 datasets respectively. The accuracy varies as shown in figure 4.1 after quality scalable quantization on LeNet. The quantization for $\emptyset$ values of 1, 2 and 4 correspond to data points {+1,-1}, {+2,-2} and {+4,-4} respectively. The quantization levels show a direct relation with the quality of deep learning models.

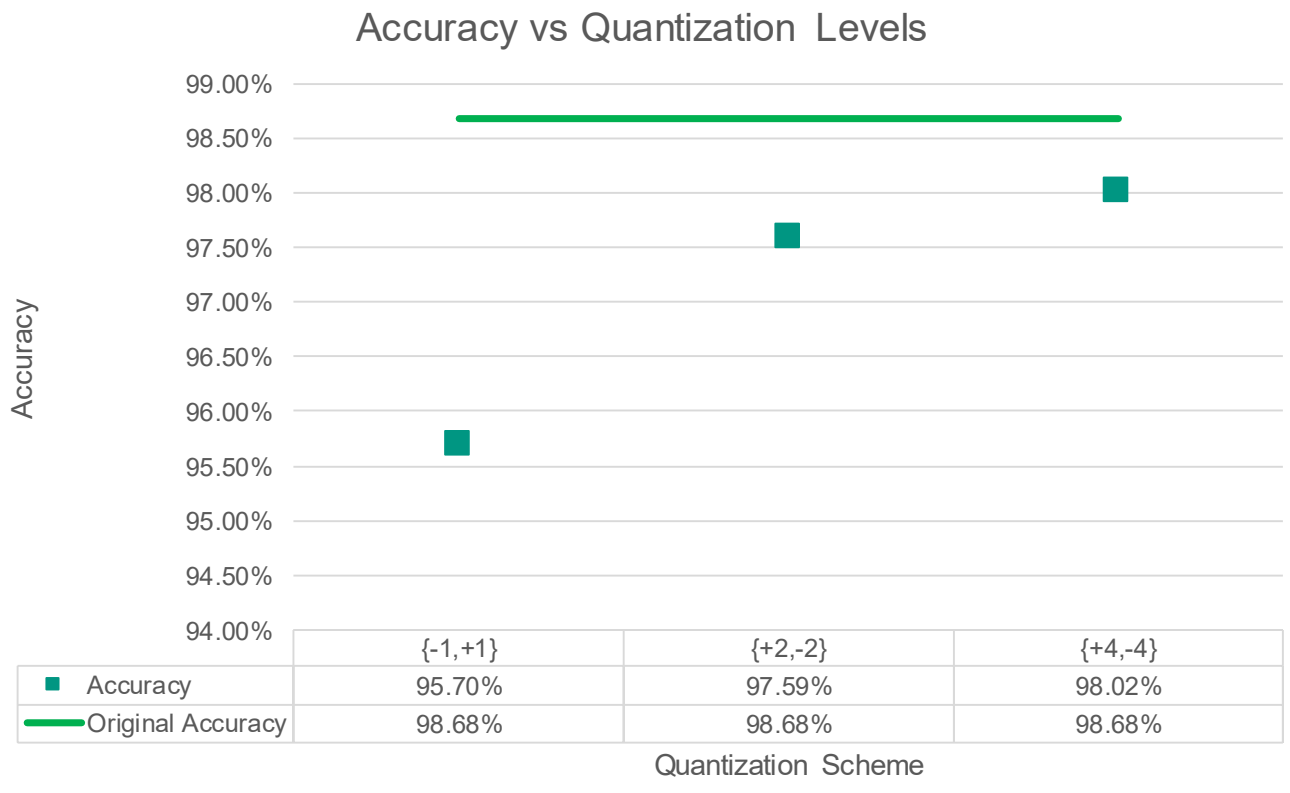

Fig. 7 Accuracy scales with quantization levels

The accuracy vs quantization graph for Cifar-10 dataset on convent architecture are shown below. Comprehensive summary of the results are shown in the bar graph and table for cifar10 dataset in Figure 7. Each four bars represent accuracy computed after quantizing 1st, 2nd, 3rd and 4th convolution layer respectively.

### C. *Design Space based on Energy Efficiency and CNN Accuracy*

The energy required for moving 32 bits from DRAM to computing chip which can be CPU, ASIC or FPGA is given as 6400pJ [8]. We have computed how much energy is required for moving encoded filter values which are quantized and for moving original weights. For calculating number of bits in non-quantized model equation 11 is used. The equation for calculating total number of bits in ternary weights or 2 bit encoded and higher quantization levels or 3 bit encoded are given by equation 12.

$$NBits_i = (FPB \times H_i \times W_i \times C_i \times Num_i) \quad (11)$$

$$NBits_{i(imp)} = (BE \times H_i \times W_i \times C_i \times Num_i) + (H_i \times W_i \times C_i \times FPB) \quad (12)$$

Here *H* and *W* represent the height and width of $i^{th}$ convolution layer weights, Num represents number of filters, *C* represents number of channels and FPB means Full Precision Bits which in our case is assumed to be 32. For various vector lengths (*N*), we demonstrated that energy efficiency for 2 bit coded although is higher than 3 bit coded models, still there remains a significant difference of accuracies between these two types of encoding schemes.

The number of bits required to save the ternary weights is 2, so 2 Bit-Encoding (BE) is sufficient for representing ternary weights and 3 Bit-Encoding (BE) is required to represent the filter weights for Ø up to 4. In Figure 9 we have plotted various design points with respect to accuracy and energy savings computed from the method we mentioned earlier. These design points are computed for varying vector lengths *N* of 2, 4, 8, 16, 32 and 64 values. The energy efficiency for ternary weights in terms of energy savings is slightly more than the higher quantization models, however this slight advantage has a much higher cost in terms of quality.

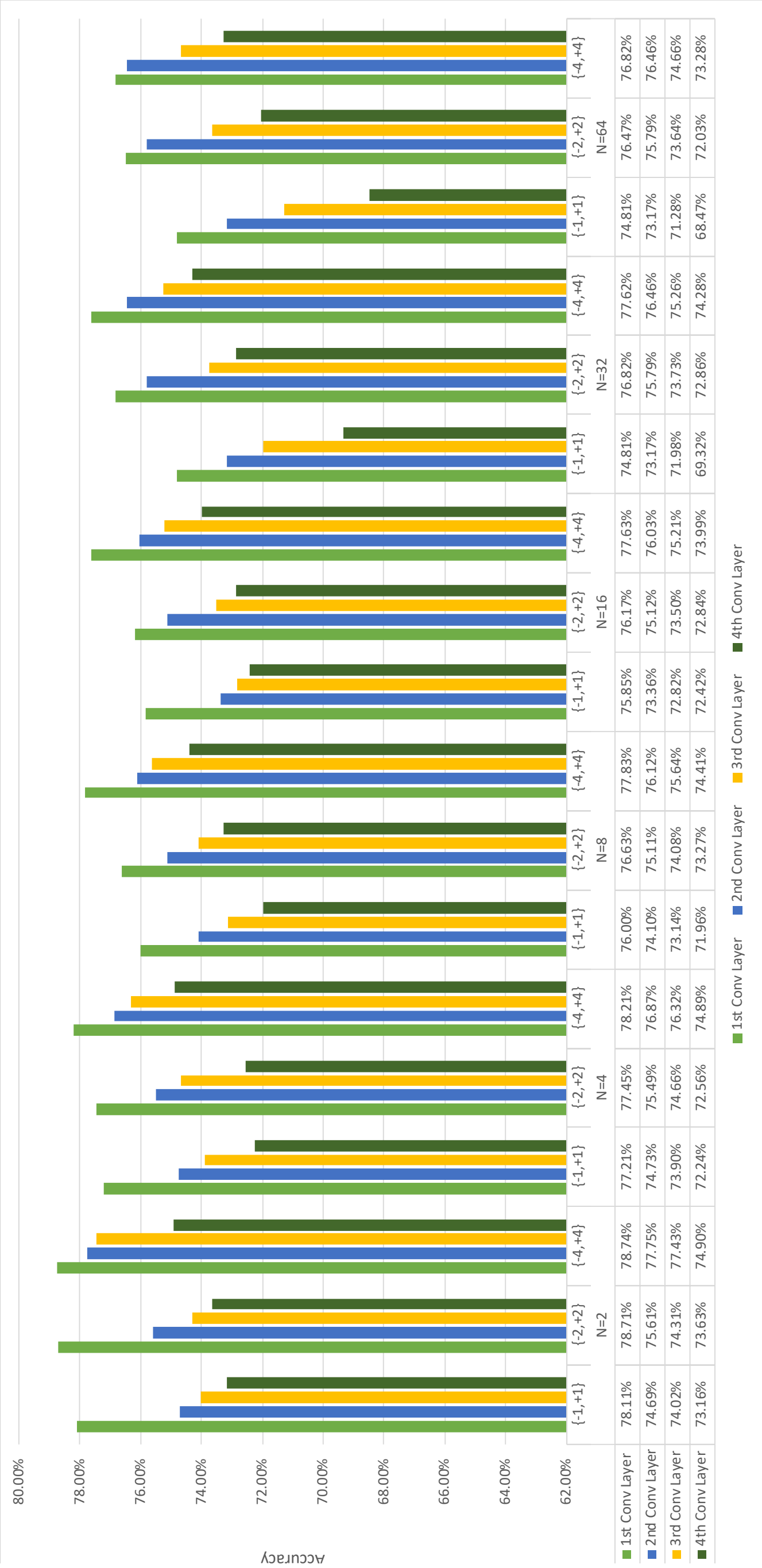

Fig. 8 ConvNet 4 layer quality scalable quantization for varying vector lengths (N)

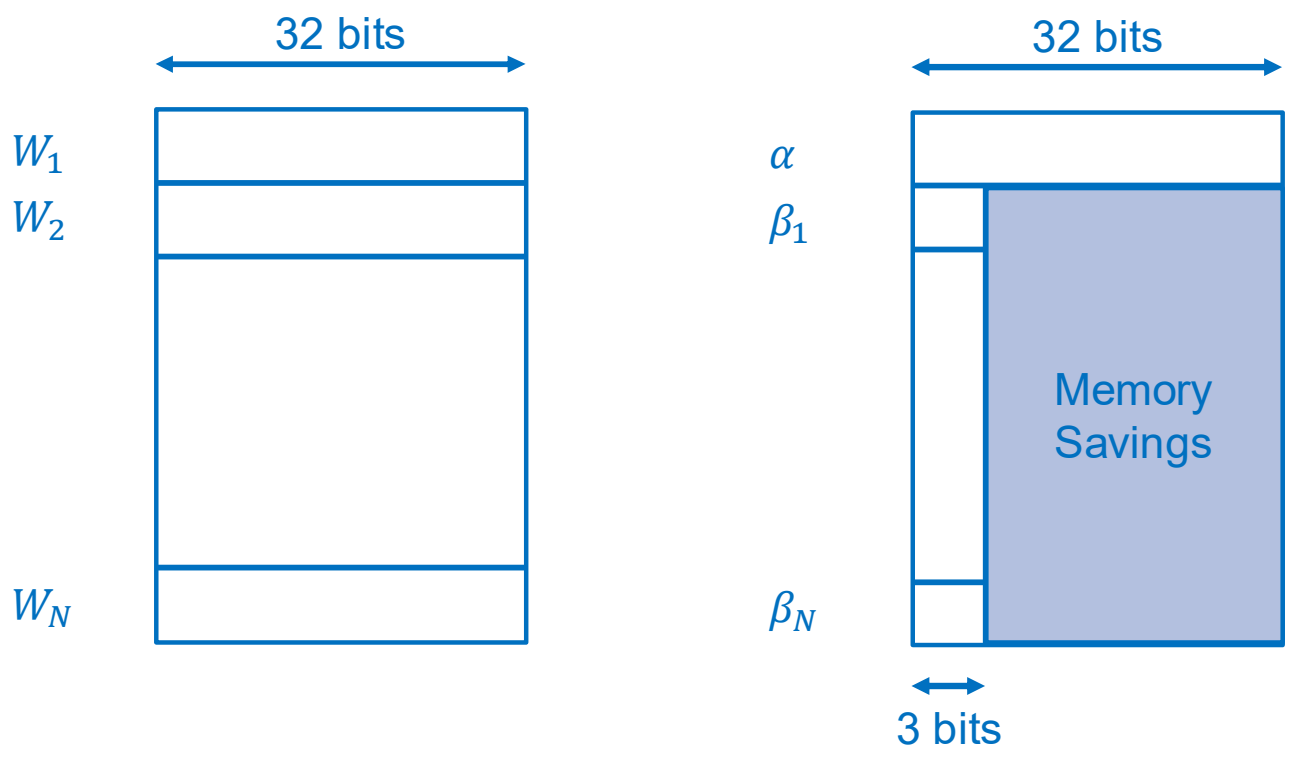

Fig. 9 Memory Savings by Encoding Vector consisting of full precision weights

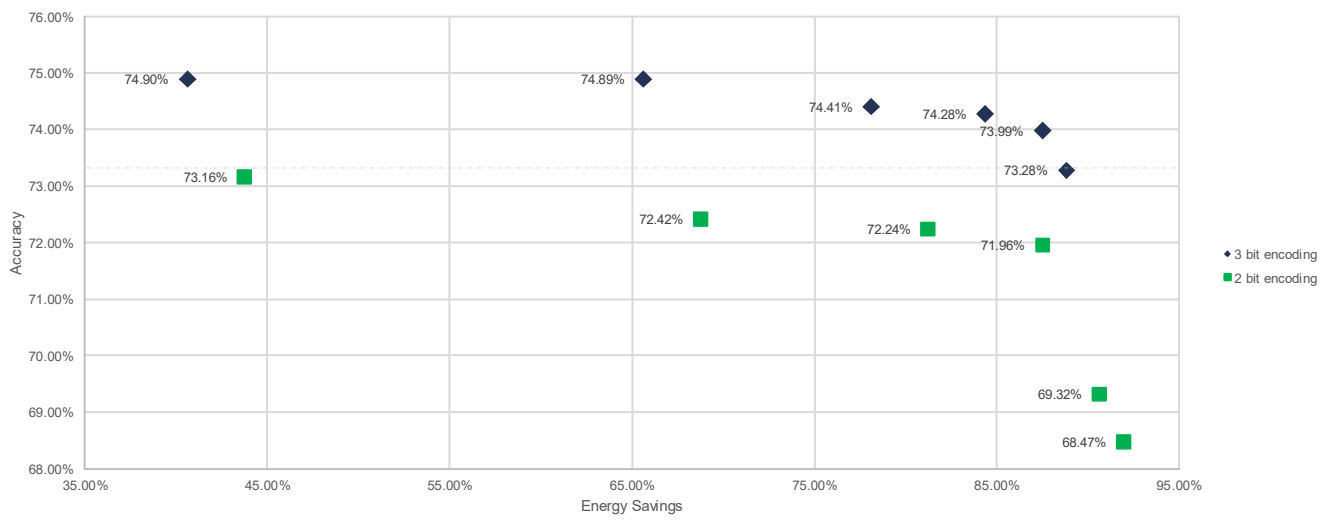

Fig. 10 Design space exploration of energy savings vs accuracy for different Vector lengths (N) and 2 and 3 bit encoding

## V. Discussion

In this section we discuss the usability of quantization methodology in terms of hardware optimization and discuss how quality scalability can be introduced at microarchitecture level using approximate multipliers.

### A. Compression of CNN Model Parameters

There can be many advantages of quantized deep learning models. For example in quantized CNN models, the weights which are quantized can be related to shifted or inverted versions of a single scalar. Therefore only a single scalar is required to be fetched from main memory along with the information about how much left or right shifting is required. The reduction in DNN model size depends on the number of channels or depth of each convolution layer, the more depth results in more values in a single filter vector. Which in turn reduces the number of bits to represent filter weights. Following figure shows reduction in the model size after quantization. Here N represents the depth of the filter and W represents a single value in filter.

### B. Quality Scalable Multipliers

The quality scalable methodology can also be complemented by using approximate multiplier technique on microarchitecture level. Since quality scalable design methodology is targeted towards edge computing devices, low power multipliers can achieve high energy savings also with quality scalable properties. We can convert the binary data in Canonic Sign Digit (CSD) format and further increase the benefit in quality scalable design. The Canonic Sign Digit (CSD) representation reduces the number of non-zeros in any binary number and thus if the multiplier is replaced with its CSD it results in less number of partial products. A quality scalable multiplier can reduce the number of partial products in the CSD representation to provide with an approximate answer with a trade off with energy consumption.

The Deep Learning model contain convolution layers which require immense multiply and accumulate operations. However because of resilience of these models to slight error or noise [24] reducing the least significant bits from the multiplier would not result in significant reduction in classification accuracy. This claim is supported by Fig. 10 below which shows that few number of non-zeros are required to accurately represent most of the values in AlexNet model filters.

These statistics were obtained using MATLAB's fi library for open source trained architecture of AlexNet CNN. Furthermore if we use CSD representation of weights we can further reduce the number of non-zero bits for representation of values of filter weights. If the number of non-zeros are restricted for quality scalable architecture design, we can reduce number of partial products generated in multiplications and without much quality degradation and we can achieve inference at very low-power.

## VI. Conclusion

In conclusion we proposed a Systematic Quality Scalable Design Methodology for mobile devices to perform machine learning operations in edge computing. The design methodology consists of Quality Scalable Quantization Scheme at a higher abstraction level. We reduced model size using quantization across filters and channels with which we achieved up to 82.49% reduction in model size for LeNet with almost same accuracy as the original DNN. For 4 layer ConvNet with 2 bit encoding we achieved 91.95% energy efficiency compared with the original model in terms of DRAM accesses with 68.47% classification accuracy. 3 bit encoding was able to achieve 73.28% accuracy with 88.82% energy efficiency with respect to original CNN. We demonstrated that increased quantization methodology despite requiring 3 bit encoding as opposed to 2 bit encoding for ternary quantization, gave far better accuracies which provided a good energy saving to accuracy ratio overall. We have also shown a general trend how quantization levels can scale quality of deep learning models.

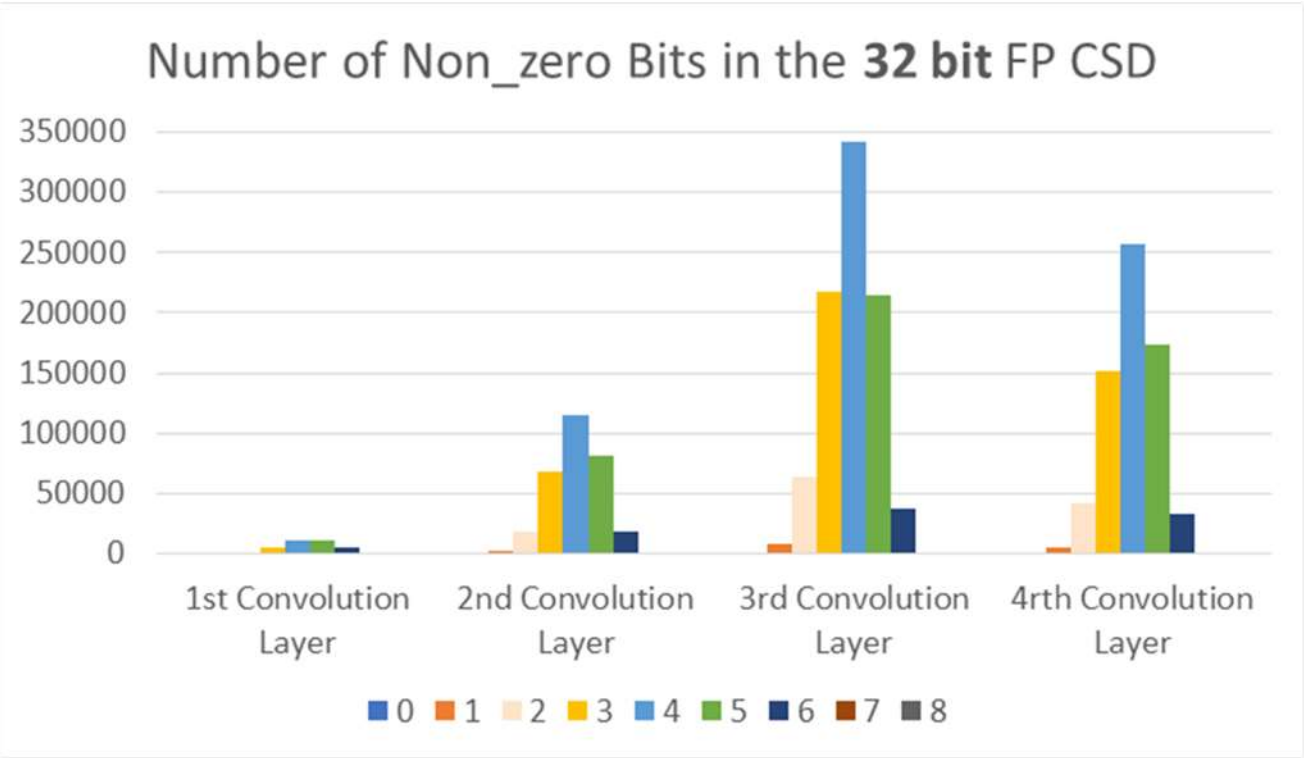

Fig. 11 Distribution of Non-Zeros in AlexNet Architecture